\documentclass[sigconf]{acmart}
\usepackage{multirow}
\usepackage{multicol}

\usepackage{subcaption}
\usepackage[flushleft]{threeparttable}
\usepackage{flushend}

\usepackage{bm}

\usepackage{CJKutf8}
\newcommand{\Chi}[2]{%
  \csname CJK*\endcsname{UTF8}{gbsn}%
    \CJKchar{"#1}{"#2}%
  \csname endCJK*\endcsname
}
\newcommand{\Jpn}[2]{%
  \csname CJK*\endcsname{UTF8}{ipxm}%
    \CJKchar{"#1}{"#2}%
  \csname endCJK*\endcsname
}
\AtBeginDocument{%
  \providecommand\BibTeX{{%
    \normalfont B\kern-0.5em{\scshape i\kern-0.25em b}\kern-0.8em\TeX}
    }
}

\copyrightyear{2021} 
\acmYear{2021} 
\setcopyright{acmcopyright}
\acmConference[KDD '21]{Proceedings of the 27th ACM SIGKDD Conference on Knowledge Discovery and Data Mining}{August 14--18, 2021}{Virtual Event, Singapore}
\acmBooktitle{Proceedings of the 27th ACM SIGKDD Conference on Knowledge Discovery and Data Mining (KDD '21), August 14--18, 2021, Virtual Event, Singapore}
\acmPrice{15.00}
\acmDOI{10.1145/3447548.3467059}
\acmISBN{978-1-4503-8332-5/21/08}

\begin{document}
% \fancyhead{}

%%
%% The "title" command has an optional parameter,
%% allowing the author to define a "short title" to be used in page headers.
\title{HGAMN: Heterogeneous Graph Attention Matching Network for Multilingual POI Retrieval at Baidu Maps}

%%
%% The "author" command and its associated commands are used to define
%% the authors and their affiliations.
%% Of note is the shared affiliation of the first two authors, and the
%% "authornote" and "authornotemark" commands
%% used to denote shared contribution to the research.

\author{
	Jizhou Huang, 
	Haifeng Wang,
	Yibo Sun, 
	Miao Fan, 
	Zhengjie Huang, 
	Chunyuan Yuan,
    \texorpdfstring{Yawen Li$^{\dagger}$}{Yawen Li}
}

\affiliation{
  \institution{Baidu Inc., Beijing, China\;\;
  $^{\dagger}$Beijing University of Posts and Telecommunications}
}

\email{{huangjizhou01, wanghaifeng, sunyibo, fanmiao, huangzhengjie, yuanchunyuan}@baidu.com;warmly0716@bupt.edu.cn}

%%
%% By default, the full list of authors will be used in the page
%% headers. Often, this list is too long, and will overlap
%% other information printed in the page headers. This command allows
%% the author to define a more concise list
%% of authors' names for this purpose.
\renewcommand{\shortauthors}{Jizhou Huang et al.}

\begin{abstract}
The increasing interest in international travel has raised the demand of retrieving point of interests (POIs) in multiple languages. This is even superior to find local venues such as restaurants and scenic spots in unfamiliar languages when traveling abroad. Multilingual POI retrieval, enabling users to find desired POIs in a demanded language using queries in numerous languages, has become an indispensable feature of today's global map applications such as Baidu Maps. This task is non-trivial because of two key challenges: (1) visiting sparsity and (2) multilingual query-POI matching. To this end, we propose a \textbf{H}eterogeneous \textbf{G}raph \textbf{A}ttention \textbf{M}atching \textbf{N}etwork (\textbf{HGAMN}) to concurrently address both challenges. Specifically, we construct a heterogeneous graph that contains two types of nodes: POI node and query node using the search logs of Baidu Maps. First, to alleviate challenge \#1, we construct edges between different POI nodes to link the low-frequency POIs with the high-frequency ones, which enables the transfer of knowledge from the latter to the former. Second, to mitigate challenge \#2, we construct edges between POI and query nodes based on the co-occurrences between queries and POIs, where queries in different languages and formulations can be aggregated for individual POIs. Moreover, we develop an attention-based network to jointly learn node representations of the heterogeneous graph and further design a cross-attention module to fuse the representations of both types of nodes for query-POI relevance scoring. In this way, the relevance ranking between multilingual queries and POIs with different popularity can be better handled. Extensive experiments conducted on large-scale real-world datasets from Baidu Maps demonstrate the superiority and effectiveness of HGAMN. In addition, HGAMN has already been deployed in production at Baidu Maps, and it successfully keeps serving hundreds of millions of requests every day. Compared with the previously deployed model, HGAMN achieves significant performance improvement, which confirms that HGAMN is a practical and robust solution for large-scale real-world multilingual POI retrieval service.
\end{abstract}

%%
%% The code below is generated by the tool at http://dl.acm.org/ccs.cfm.
%% Please copy and paste the code instead of the example below.
%%




%%
%% This command processes the author and affiliation and title
%% information and builds the first part of the formatted document.
\maketitle

% {\fontsize{8pt}{8pt} \selectfont\textbf{ACM Reference Format:}\\
% Jizhou Huang, Haifeng Wang, Yibo Sun, Miao Fan, Zhengjie Huang, Chunyuan Yuan, Yawen Li. 2021. HGAMN: Heterogeneous Graph Attention Matching Network for Multilingual POI Retrieval at Baidu Maps. In \textit{Proceedings of the 27th ACM SIGKDD Conference on Knowledge Discovery and Data Mining (KDD '21), August 14--18, 2021, Virtual Event, Singapore.} ACM, New York, NY, USA, 9 pages. \url{https://doi.org/10.1145/3447548.3467059}} 

\section{Introduction}

\begin{figure}[t!]
	\centering
	\includegraphics[width=1.0\linewidth,trim={0.2cm 0.2cm 0.2cm 0.03cm},clip]{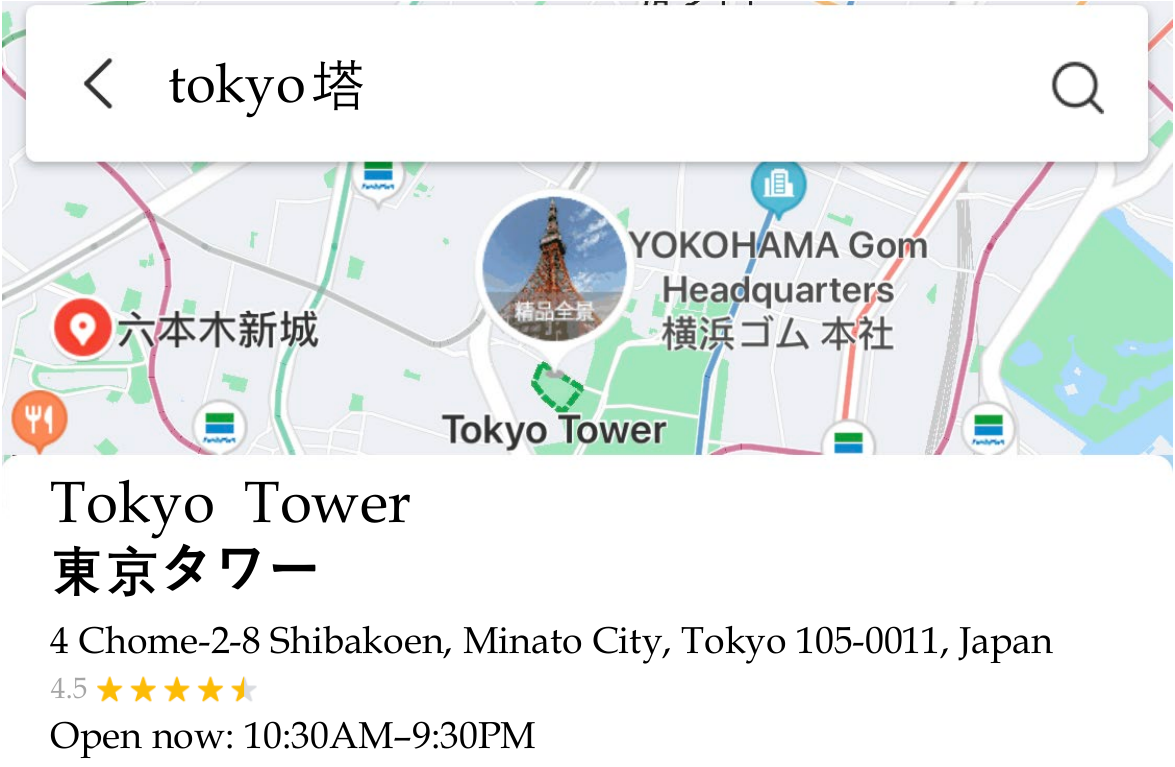}
	\caption{An example of the multilingual POI search result for the query ``tokyo \Chi{58}{54}'' at Baidu Maps. In this case, the multilingual POI retrieval module helped the user obtain the desired POI named with English words ``Tokyo Tower'' or Japanese words ``\Jpn{67}{71}\Jpn{4E}{AC}\Jpn{30}{BF}\Jpn{30}{EF}\Jpn{30}{FC}'' by using a query consisting of mixed-language characters, i.e., an English word ``Tokyo'' and a Chinese character ``\Chi{58}{54}''.}
	\label{fig:mulretexample}
\end{figure}

As one of the key components of the search engines in almost all global map applications, such as Baidu Maps, multilingual POI retrieval plays a significant role in providing on-demand map services as the retrieved results directly influence the success or failure of routing and navigation, and hence impact the long-term user experience. For the 169 million Chinese tourists who traveled abroad in 2019 \cite{chinastas2020}, Baidu Maps, which covers more than 150 million POIs in 200 countries and territories worldwide, is their prior choice to find specific locations and navigate to desired destinations. Figure~\ref{fig:mulretexample} shows an example of the multilingual POI retrieval feature at Baidu Maps, where the query ``Tokyo\Chi{58}{54}'' consists of an English word ``Tokyo'' and a Chinese character ``\Chi{58}{54}'', meanwhile the name of the retrieved POI is composed of English words ``Tokyo Tower'' or Japanese words ``\Jpn{67}{71}\Jpn{4E}{AC}\Jpn{30}{BF}\Jpn{30}{EF}\Jpn{30}{FC}''. To enable users who are traveling abroad to obtain their desired POIs effectively when finding local venues in unfamiliar languages and areas, it is crucial for a multilingual POI retriever to fill the gaps between queries and POIs in different languages and formulations. 

To build an effective multilingual POI retriever in both the academic and industrial fields, we must address two key challenges: 
\begin{itemize}
	\item \textbf{Visiting Sparsity.}	
	To the best of our knowledge, existing approaches on multilingual POI retrieval for industrial use mainly leverage large-scale user click logs for query-POI relevance scoring.
	However, the average visits of 150 million POIs at Baidu Maps are highly sparse.
	We empirically study a large-scale search log of Baidu Maps, which contains billions of search records. Statistics show that only 6.4\% of the POIs have been clicked by one or more users. 
	The effectiveness of a POI retrieval model would significantly decline when handling the majority of POIs that have sparse click logs.
	
	\item \textbf{Multilingual Query-POI Matching.} In real applications, most of the users search the overseas POIs by their native languages, which are more likely to be inconsistent with the languages of the target POIs. For example, a Chinese user may search the ``Tokyo Tower'' located in Japan using queries composed of Chinese words, meanwhile, the information of this POI is probably maintained in Japanese or English. As a result, a simple literal matching method cannot meet the demands of such cross-language retrieval. Moreover, queries are sometimes mixed keyboard inputs of multi-languages (e.g., English and Japanese, Chinese and Pinyin Alphabets), which further necessitates multilingual POI retrieval. 
\end{itemize}

In this paper, we present our recent efforts in designing and implementing an effective multilingual POI retrieval framework, which has already been deployed in production at Baidu Maps and has achieved great success in addressing both problems, as illustrated by Figure~\ref{fig:mulretexample}.
The framework can provide a data sparsity-tolerant multilingual POI retrieval function, which facilitates tens of millions of users to find their desired POIs every day.

This new framework is powered by a \textbf{H}eterogeneous \textbf{G}raph \textbf{A}ttention \textbf{M}atching \textbf{N}etwork (\textbf{HGAMN}).
Specifically, we first construct a heterogeneous graph that contains two types of nodes: POI node and query node using the search logs of Baidu Maps. 
In this graph, to alleviate the visiting sparsity problem, we construct edges between different POI nodes to 
link the low-frequency POIs with the high-frequency ones, which enables the transfer of knowledge from the latter to the former.
To address the multilingual query-POI matching challenge, we construct edges between POI and query nodes based on the co-occurrences between queries and POIs, where queries in different languages and formulations can be aggregated for individual POIs.
Upon the constructed graph, we design an attention-based network to learn the representations of POI and query nodes.
Then, we use a multi-source information learning module to learn the location and multilingual text representations of the queries and POIs.
Finally, we fuse the node representations of a POI and its linked queries via a cross-attention module and use the fused representation to calculate the relevance score between the user's query and a candidate POI.
To facilitate the model training, we apply an in-batch negative sampling strategy~\cite{karpukhin2020dense} to produce more sample pairs and increase the number of training examples.

We evaluate HGAMN both offline and online using large-scale real-world datasets. For offline evaluation, the training and test sets consist of tens of millions of search records for several months, covering hundreds of cities and tens of millions of POIs worldwide. Experimental results show that HGAMN achieves substantial (absolute) improvements compared with several mainstream methods. For online evaluation, we launch our framework online to serve a portion of the search traffic at Baidu Maps. A/B testing was conducted between HGAMN and the previously deployed models. Experimental results show that the improvements are consistent with those obtained by the offline evaluation.

The main contributions can be summarized as follows:
\begin{itemize}
	\item \textbf{Potential impact:} We propose an end-to-end neural framework, named HGAMN, as an industrial solution to the multilingual POI retrieval task in global map applications. In addition, this framework has already been deployed in production at Baidu Maps, and it successfully keeps serving hundreds of millions of POI search requests every day.

	\item \textbf{Novelty:} The design of HGAMN is driven by the novel idea that addresses the data sparsity problem and the multilingual matching problem by enhancing the representations of POIs via a heterogeneous graph. 
	
	\item \textbf{Technical quality:} We evaluate HGAMN both offline and online using large-scale real-world datasets. Extensive experimental results show that our framework achieves significant improvements on multiple evaluation metrics compared with several mainstream methods.
	
	\item \textbf{Reproducibility}: 
    We have made the source code publicly available  at~\url{https://github.com/PaddlePaddle/Research/tree/master/ST_DM/KDD2021-HGAMN/}. 
\end{itemize}

\begin{figure*}[ht]
	\centering
	\includegraphics[width=\textwidth,clip]{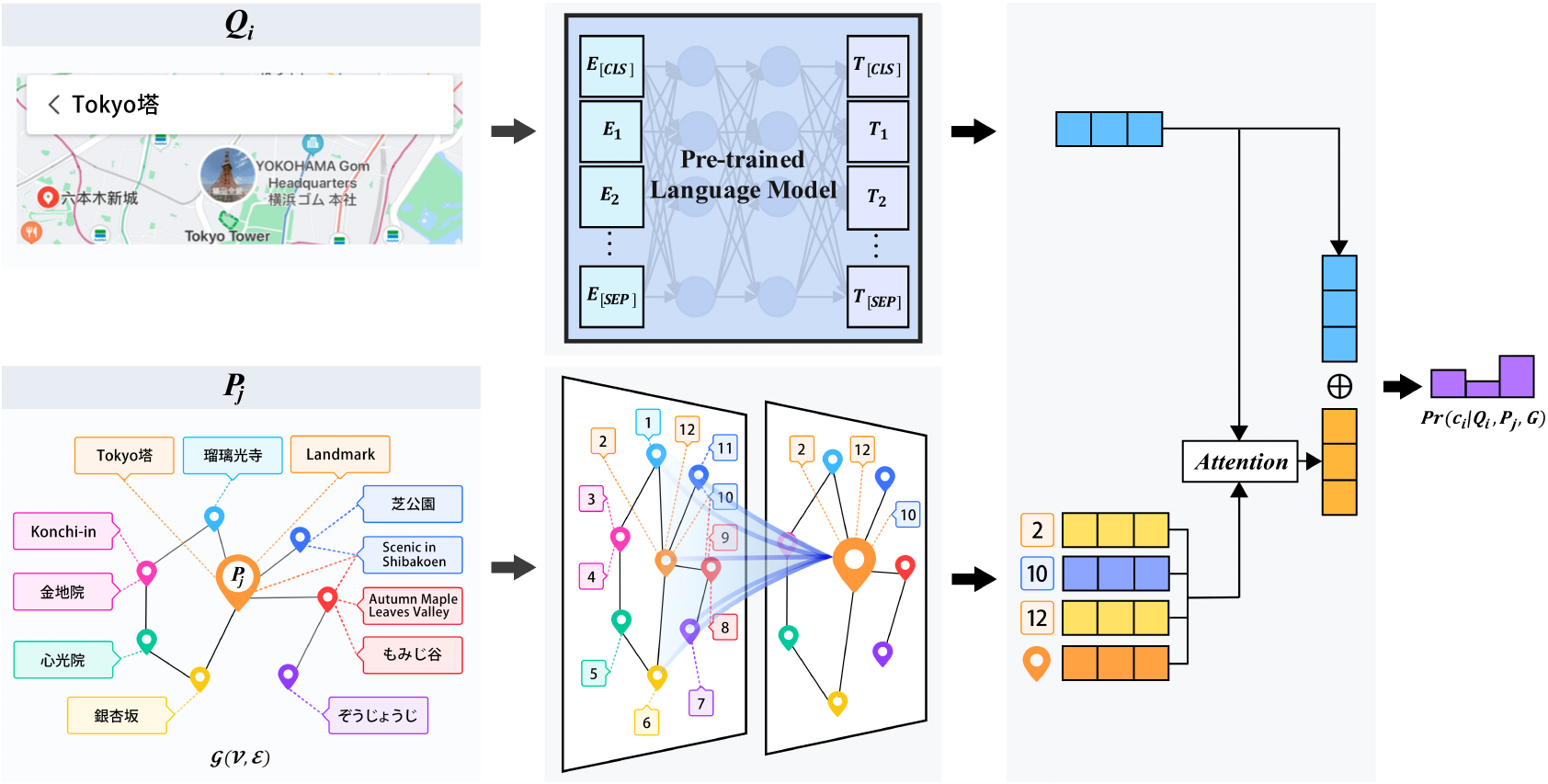}
	\caption{
		The architecture of HGAMN.
	}
	\label{fig:framwork}
\end{figure*}

\section{HGAMN}
\label{sec:framework}
HGAMN consists of three modules: multi-source information learning module, heterogeneous graph learning module, and POI ranker module. First, we feed a query, the candidate POIs, and the historical queries to the multi-source information learning module to learn the text and location representations of them. Then, we construct the heterogeneous graph of different POIs and historical queries. The constructed graph enables us to learn the POI representations from it by the heterogeneous graph learning module. Finally, we calculate the relevance score between the representations of the query and the candidate POIs by the POI ranker module. Figure~\ref{fig:framwork} shows the architecture of HGAMN. Subsequently, we introduce them in detail.

\subsection{Multi-Source Information Learning}  \label{multi_source_info}
Unlike traditional text retrieval, POI retrieval in map services mainly measures the relevance between a query and POIs rather than plain text. Besides its name, a typical POI also contains other multi-sourced and heterogeneous information such as the address, category, and GPS coordinates. Utilizing such information can facilitate retrieving more relevant POIs. Here we mainly introduce the location and text representations of a query $q \in \mathcal{Q}$ and a POI $P_i \in \mathcal{P}$, where $\mathcal{Q}$ and $\mathcal{P}$ denote a set of query and POI, respectively.

\subsubsection{GPS Encoding}
\label{gpsencoding}
POI's GPS coordinates are numerical pairs consisting of longitude and latitude. However, in the online system, the coordinates are usually stored as a Geohash string for its better properties: (1) it is easy to be used to index the POI and (2) it is convenient to be used to calculate the distance of two POIs. 

Instead of directly taking this numerical pair as a 2-dimensional feature vector, we use the Geohash algorithm~\cite{morton1966computer} to encode the geographic coordinates into a short string of letters and digits. Specifically, given the latitude $x_v$ and longitude $y_v$ of a POI, the Geohash algorithm is performed as follows:
\begin{equation}
s_{GPS} = \mathbf{Geohash}((x_v, y_v))  \,,
\end{equation}
where the length $|s_{GPS}| \in [1, 12]$. 

Given the Geohash string $s_{GPS}=$``wx4g09np9p'', we split the string to character sequence and add `[PAD]' at the beginning of the sequence if its length is less than 12, i.e., $X = [$`[PAD]', `[PAD]', `w', `x', `4', `g', `0', `9', `n', `p', `9', `p' $]$. Then, we transform them into character embeddings $\mathbf{X} \in \mathbb{R}^{12 \times d_c}$, where $d_c = 64$ is the dimension of the character embedding.

An essential property of Geohash string is that POIs with a longer common prefix are closer to each other in geographic distance. Thus, the Geohash string is order-sensitive. To encode this kind of property, we utilize the bidirectional gated recurrent unit to encode the character embeddings, which is formulated by:
\begin{equation}
\mathbf{\widetilde{h}}_t = [\overrightarrow{\mathbf{GRU}}(\mathbf{X}_t); \overleftarrow{\mathbf{GRU}}(\mathbf{X}_t)]  \,. \\
\end{equation}

The last state $\mathbf{\widetilde{h}}_{12}$ is used as the representation of the POI's GPS. We use this module to transform each POI's GPS coordinates into an embedding, and obtain an embedding matrix $\mathbf{G} \in \mathbb{R}^{|\mathcal{P}| \times d}$, where $|\mathcal{P}|$ denotes the size of $\mathcal{P}$. 

We regard a query's location as the place where the user is typing in the query. Similarly, we can obtain the query's location representation $\mathbf{G}_u$ according to the user's GPS coordinates.

\begin{figure*}[ht]
	\centering
	\includegraphics[width=\textwidth,clip]{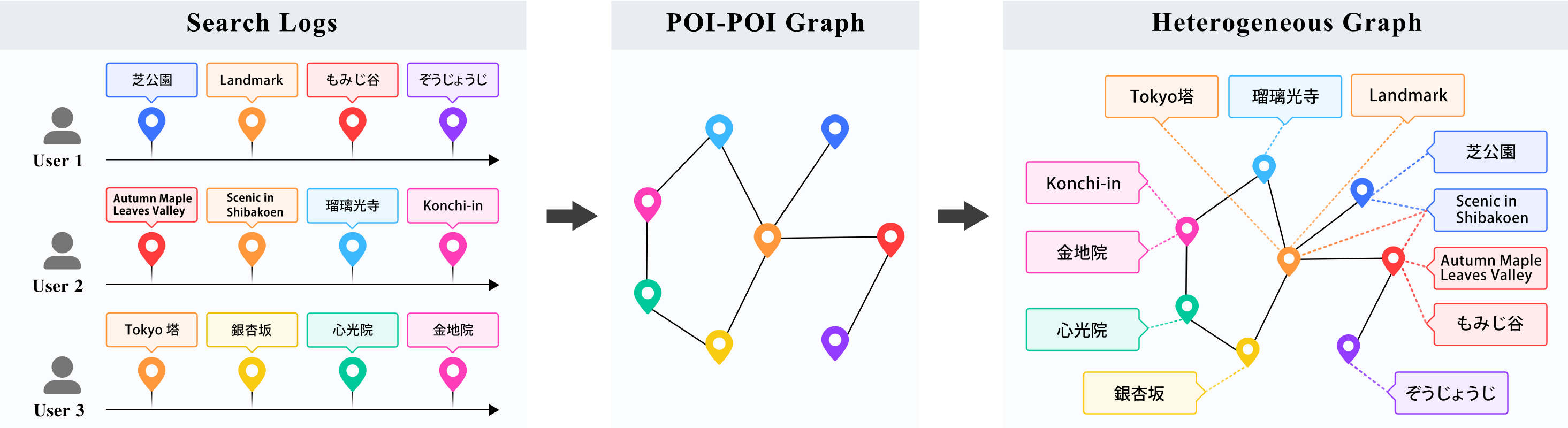}
	\caption{
		The process of constructing the heterogeneous graph $\mathcal{G}$. 
	}
	\label{fig:construct_graph}
\end{figure*}

\subsubsection{Text Encoding}  \label{text_encoder}
For multilingual POI retrieval in map services, the text data such as queries, POI names, and POI addresses are critical for improving the retrieval performance.

To better handle the multilingual matching problem, we take a sequence (such as a query or POI name) consisting of multilingual characters and alphabets as input and adopt a pre-trained language model to obtain its representation.

Specifically, we use the pre-trained language model ERNIE~\cite{sun2019ernie} as the basic component, which shows better performance on extracting multilingual features and semantic information. 
We directly utilize ERNIE to obtain $q$'s text representation $\mathbf{q}$ by:
\begin{equation}
	\mathbf{q} = \mathbf{ERNIE}([c_1, c_2, \ldots, c_L]) \,,  \\
\end{equation}
where $[c_1, c_2, \ldots, c_L]$ is the character sequence of the query.

After analyzing the query logs, we found that a query's location helps retrieve the desired POI because users usually demand the nearest target. To utilize such location features, we combine a query's location representation with its text representation. Thus, the final representation of a query is represented as:  $\widetilde{\mathbf{q}} = \mathbf{q} + \mathbf{G}_u$.

For each POI $P_i$, we use $\mathbf{Q}_{P_i}$ to denote the matrix of the representations of the top-4 queries associated with it, i.e., $\mathbf{Q}_{P_i} = [\widetilde{\mathbf{q}}_1, \widetilde{\mathbf{q}}_2, \widetilde{\mathbf{q}}_3, \widetilde{\mathbf{q}}_4]$.

For each POI $P_i$, we concatenate its name and address as a long character sequence and apply ERNIE to extract its text representation. Similarly, we combine $P_i$'s location and text representations to obtain its final representation $\mathbf{P}_i \in \mathbb{R}^{1 \times d}$ by:
\begin{equation}
	\mathbf{P}_i = \mathbf{ERNIE}([x_1, x_2, \ldots, x_L, a_1, a_2, \ldots, a_L]) + \mathbf{G}_i \,,  \\
\end{equation}
where $[x_1, x_2, \ldots, x_L]$ is the character sequence of $P_i$'s name and $[a_1, a_2, \ldots, a_L]$ is the character sequence of $P_i$'s address. $\mathbf{G}_i$ is the GPS encoding of $P_i$. We stack each POI's embedding together as a POI embedding matrix $\mathbf{P} \in \mathbb{R}^{|\mathcal{P}| \times d}$.

\subsection{Heterogeneous Graph Learning} \label{graph_learning}
Here, we introduce how to construct the heterogeneous graph from search logs and how to learn POI representations from the heterogeneous graph.

\subsubsection{Graph Construction}
Multilingual POI retrieval faces the visiting sparsity and multilingual matching problems. To alleviate the visiting sparsity problem, we construct edges between different POI nodes to link the low-frequency POIs with the high-frequency ones, which enables the transfer of knowledge from the latter to the former.
Furthermore, to mitigate the multilingual matching problem, we construct edges between POI and query nodes based on the co-occurrences between queries and POIs. Thus, the graph can aggregate queries in different languages and formulations for each POI. 
As shown by Figure~\ref{fig:construct_graph}, both types of nodes and edges constitute a heterogeneous graph $\mathcal{G}(\mathcal{V}, \mathcal{E})$. 

The construction of the heterogeneous graph is as follows. A user's search behaviors produce a visited POI sequence in search logs. We extract relations of POI-POI from the historical search sequences. A search sequence is a period of time that consists of ``a sequence of interactions'' for the similar information need~\cite{10.1145/3396501}, which can reflect the similarities of successive POIs. To capture the similarity between two POIs, we define the co-occurrence frequency of them in the search sequences as the graph's edge.
To extract this kind of relation, a 2-gram sliding window is perform on the search sequences. We apply the pointwise mutual information (PMI) to calculate the weight of the edges by:
\begin{equation}
\begin{split}
& \mathbf{A}^{pp}_{ij} = \text{PMI}(P_i, P_j) = log \frac{Pr(P_i, P_j)}{Pr(P_i) \cdot Pr(P_j)}  \,,  \\
& Pr(P_i, P_j) = \frac{\#W(P_i, P_j)}{\#W}  \,,  \\
& Pr(P_i) = \frac{\#W(P_i)}{\#W}  \,,  \\
\end{split}
\end{equation}
where $\#W(P_i, P_j)$ is the number of sliding windows that contain both $P_i$ and  $P_j$. $\#W$ denotes the number of sliding windows, and $\#W(P_i)$ is the number of sliding windows that contain $P_i$. 

After typing in a query, a user would click on the desired POI from a list of ranked POIs that the POI search engine suggested.
This process produces a large-scale query-POI pairs where the multilingual expressions of each POI can not only effectively mitigate the multilingual matching problem, but also bridge the semantic gap between queries and POIs. For example, users usually make spelling errors or use abbreviations, which would lead to poor results when directly matching query and POI text information. Motivated by this observation, we try to model the relations between historical queries and POIs. 

Specifically, we select the top-4 searched queries for each POI and connect an edge for every POI and its historical query nodes for POI-Query relations. In this way, we can build connections between POIs and Queries. Formally, the adjacency matrix can be formulated as follows:
\begin{equation}
\begin{split}
\mathbf{A}^{pq}_{ij} = \frac{c_{i,j}}{\sum_{k=1}^{|\mathcal{Q}_{P_i}|} c_{i,k}}  \,,
\end{split}
\end{equation}
where $c_{i,j}$ is the frequency of query-POI pair $(q_j, P_i)$, $q_j \in \mathcal{Q}_{P_i}$.

\subsubsection{Heterogeneous Graph Learning}  \label{poi_query_graph_learning}
To learn representations of POIs and queries from the heterogeneous graph, we use attention-based graph neural network to aggregate neighbors for generating a distributed representation of each node based on the heterogeneous graph, which enables us  to learn a high-level hidden representation for each vertex.

As shown by Figure~\ref{fig:construct_graph}, there are two types of nodes (POI node and query node) and two types of edges (POI-POI correlation and POI-Query semantic relation) in the graph. For a certain node $n_i$, its initial embedding: $\mathbf{n}_i \in \mathbb{R}^{d_n}$, is randomly initialized by a uniform distribution.

First, we introduce how to produce edge embeddings for the heterogeneous graph. We apply an aggregator proposed in GraphSAGE~\cite{hamilton2017inductive} to integrate neighbor node embeddings for edge embedding $\mathbf{e}_j$:
\begin{equation}
    \mathbf{e}^{(k)}_j = \sigma( \mathbf{\max}(\{ \mathbf{W}^{(k)} \mathbf{n}_{t}, n_{t} \in \mathcal{N}_j \} ) )  \,,
\end{equation}
where $\mathbf{W}^{(k)}$ is the trainable weight of $k$-th layer, $\sigma(\cdot)$ is the sigmoid activation function. $\mathcal{N}_j$ is the neighbor set of edge $e_j$.

Suppose the node $n_i$ has $m$ edges connected with it, we concatenate edge embeddings for the node $n_i$ as $\mathbf{E}_i \in \mathbb{R}^{m \times d_e}$: 
\begin{equation}
\mathbf{E}_i = (\mathbf{e}_{i,1}, \mathbf{e}_{i,2}, \ldots, \mathbf{e}_{i,m}) \,.
\end{equation}

Next, we apply the cross attention mechanism to fuse $\mathbf{E}_i$ into a vector $\mathbf{\tilde{e}}_i \in \mathbb{R}^{d_e}$ by: 
\begin{equation}
\begin{split}
    \bm{\alpha}_{i} &= \mathbf{softmax}(\mathbf{n}_i \tanh(\mathbf{W}_r \mathbf{E}^T_i) \odot \mathbf{A}^{(r)}_{i} )  \,, \\
    \mathbf{\tilde{e}}_i &= \bm{\alpha}_i^{T}\mathbf{E}_i \,, \\
\end{split}
\end{equation}
where $\odot$ denotes the element-wise multiplication operation. $\bm{\alpha}_{i} \in \mathbb{R}^{m}$ is the coefficients. $\mathbf{W}_r \in \mathbb{R}^{d_n \times d_e}$ is a trainable weight for edge type $r$. $\mathbf{A}^{(r)}_{i} \in \{\mathbf{A}_{i,1:m}^{pp}, \mathbf{A}_{i,1:m}^{pq} \}$ denotes the weights of $m$ edges which belong to the type $r$ and connect with $n_i$.

Then, the overall node representation of $n_i$ can be computed by:
\begin{equation}
    \mathbf{\widetilde{n}}_i = \mathbf{n}_i + \mathbf{W}_1\mathbf{\tilde{e}}_i  + \mathbf{W}_2 \mathbf{n}^{\prime}_i  \,,
\end{equation}
where $\mathbf{W}_1 \in \mathbb{R}^{d_n \times d_e}, \mathbf{W}_2 \in \mathbb{R}^{d_n \times d}$ are two trainable parameters and $\mathbf{n}^{\prime}_i \in \{ \mathbf{Q}_{P_i},\mathbf{P}_i \}$.
$\mathbf{Q}_{P_i}$ and $\mathbf{P}_i$ denote ${P}_i$'s associated query representations and its POI embedding, respectively.

Finally, the graph learning module produces the representations of all POIs:   $\widetilde{\mathbf{P}} \in \mathbb{R}^{|\mathcal{P}| \times d_n}$, and the representations of all queries: $\widetilde{\mathbf{Q}} \in \mathbb{R}^{|\mathcal{Q}| \times d_n}$.

\subsection{POI Ranker}

\label{sec:poi_ranker}
The POI ranker module calculates the relevance between a query $q$ and a candidate POI $P_i$ based on the learned representations.
This module also considers $P_i$'s historical queries $\mathcal{Q}_{P_i}$ when predicting the relevance since $\mathcal{Q}_{P_i}$ conveys substantial evidence to bridge the semantic gap between $P_i$ and $q$.
Both $P_i$ and $\mathcal{Q}_{P_i}$ contain essential information for calculating the relevance, but their importance is different. How to automatically determine the importance of them for measuring the relevance is still a challenge.

In this paper, we apply an attention module to automatically determine their importance and fuse them as a feature vector. Specifically, we regard the representation of $q$ as the key, while regard the representations of $P_i$ and $\mathcal{Q}_{P_i}$ as the value. We stack the representations of $P_i$ and $\mathcal{Q}_{P_i}$ as a new matrix  $\mathbf{M} = [\widetilde{\mathbf{P}}_i, \widetilde{\mathbf{Q}}_{P_i} ]$. Each attention weight $\phi_k$ is defined as follows:
\begin{equation}
\begin{split}
& s_k = \mathbf{W}_4\tanh ([\mathbf{\widetilde{q}}; \mathbf{M}_k] \mathbf{W}_3 + b)  \,,  \\
& \phi_k = \frac{exp(s_k)}{\sum_{j=1}^{|\mathbf{M}|} exp(s_j) } \,,  \\
\end{split}
\end{equation}
where $\mathbf{W}_3 \in \mathbb{R}^{2d_n \times d_n}$ and  $\mathbf{W}_4 \in \mathbb{R}^{1 \times d_n}$ are trainable matrices.

We use the attention weight to fuse the representations of $P_i$ and $\mathcal{Q}_{P_i}$ by:
\begin{equation}
  \mathbf{m} = \sum_{k=1}^{|\mathbf{M}|} \mathbf{\phi}_k \mathbf{M}_k  \,, \\
\end{equation} 
where $\mathbf{m} \in \mathbb{R}^{d_n}$ is the fused POI representation.

Finally, we concatenate $\mathbf{\widetilde{q}}$ with $\mathbf{m}$, and feed them into the output softmax layer for relevance calculation by:
\begin{align}
Pr(c_{i}|q, P_i, \mathcal{G}) = \mathbf{softmax}([\mathbf{\widetilde{q}}; \mathbf{m}] \mathbf{W}_v)  \,,
\end{align}
where $\mathbf{W}_v \in \mathbb{R}^{2d_n \times 2}$ is the trainable parameter, and $Pr(c_{i}|q, P_i, \mathcal{G}) $ is the probability vector of a category $c_i \in \{0, 1\}$. The category 1 (0) indicates that $P_i$ is relevant (irrelevant) to $q$. We use the output probability of category 1 as the score for ranking.

\subsection{Model Training}
We train the model in a supervised manner by minimizing the cross-entropy loss of relevance classification described above, whose loss function is defined as follows:
\begin{equation}
\mathcal{L} = - \sum_{i=1}^{|\mathcal{P}|} y_{i} \mathbf{log} Pr(c_{i}|q, P_i, \mathcal{G})  \,,
\end{equation}
where $|\mathcal{P}|$ denotes the amount of total training POIs, and $y_{i}$ is the label of the instance POI $P_i$.

To increase the number of training instances inside each batch and improve the computing efficiency, we apply an in-batch negative sampling strategy~\cite{karpukhin2020dense}. Specifically, assuming that we have $B$ queries in a mini-batch, each one is associated with a relevant POI. Let $\mathbf{\hat{Q}}$ and $\mathbf{\hat{P}}$ be the $(B \times d)$ matrix of query and POI embeddings in a batch of size $B$. $\mathbf{S} = \mathbf{\hat{Q}} \mathbf{\hat{P}}^T$ is a $(B \times B)$ matrix of similarity scores, where each row corresponds to a query, paired with $B$ POIs. In this way, we reuse computation and effectively train on $B^2$ $(q_m, P_n)$ query-POI pairs in each batch. Any $(q_m, P_n)$ pair is a positive example when $m = n$, and negative otherwise. This procedure creates $B$ training instances in each batch, where there are $B - 1$ negative POIs for each query.

\section{Experiments}  \label{Experiments}
To thoroughly test HGAMN, we conduct extensive experiments in both offline and online settings. 

\subsection{Comparison Models}
\label{sec:model_comp}
We evaluate HGAMN against the following four groups of methods. Furthermore, to understand the relative importance of several facets of HGAMN, variations of this model with different settings are implemented for comparison.

\subsubsection{Text Matching Group}
\begin{itemize}
	\item \textbf{DSSM}~\cite{huang2013learning} is a widely-used text matching model in which a deep neural network is employed to predict the relevance between keywords and documents. In our experiments, for all DSSM based models, we treat queries as keywords while POI name and POI address as documents. 
	
	\item \textbf{ARC-I}~\cite{hu2014convolutional} uses pre-trained word embeddings to represent the text. It then uses a convolutional network to learn the semantic features and feeds the feature vectors to a multi-layer perceptron for prediction. 
	
	\item \textbf{Conv-DSSM}~\cite{shen2014latent} extends DSSM by adding extra convolutional layers to extract sentence-level features from n-gram word representations.
\end{itemize}

\subsubsection{Query-POI Matching Group}
\begin{itemize}
	\item \textbf{DPSM}~\cite{zhao2019poi} is a POI latent semantic model based on neural networks, which extracts query and POI semantic features for the similarity calculation.
	
	\item \textbf{PALM}~\cite{zhao2019incorporating} is an attention-based neural network. It uses semantic similarity and geographic correlation to quantify the query-POI relevance. 
	
\end{itemize}

\subsubsection{Our Model and Its Variants}
\begin{itemize}
	\item \textbf{HGAMN} is the complete model defined in Section~\ref{sec:framework}. In this setting, we use it independently as a POI retriever to return the desired POIs.
	
	\item \textbf{HGAMN w/o POI-POI Graph}. In this setting, we remove the edges between different POIs in the graph learning module described in Section~\ref{poi_query_graph_learning}. The removed part is designed to mitigate the visiting sparsity problem.  
	
	\item \textbf{HGAMN w/o POI-Query Graph}. In this setting, we remove the edges between different POIs and queries in the graph learning module described in Section~\ref{poi_query_graph_learning}. The removed part is designed to mitigate the multilingual matching problem.  
	
	\item \textbf{HGAMN w/o Heterogeneous Graph}. In this setting, we remove the entire graph learning module described in Section~\ref{poi_query_graph_learning} and directly use the query and POI's representations described in section~\ref{multi_source_info} for calculation.  
\end{itemize}

\subsubsection{Online Model Group}
\begin{itemize}
    \item \textbf{LTR} is the basic model for online multilingual POI retrieval system at Baidu Maps \cite{li2020personalized,huang2020personalized}. It adopts GBRank~\cite{zheng2007regression} as the specific learning-to-rank model. This model mainly uses heuristic features, including the popularity of POIs, the demographic information on users, and the spatial-temporal features of each POI, such as the frequency of search on specific types of POIs at different times and locations.
    \item \textbf{LTR + HGAMN} is trained with all the features employed by LTR and the similarity feature computed by HGAMN. It is expensive to  directly deploy HGAMN online to serve hundreds of millions of requests every day. For this reason, we instead use the feature generated by HGAMN offline as one of the features fed to the LTR model.
\end{itemize}

\subsection{Offline Evaluation}
\subsubsection{Dataset}
The services of Baidu Maps cover over 200 countries and territories worldwide, where the sessions on POI search dominate about 80\% search traffic. A POI search session refers to a sequence of interactions between a user and the POI search engine.

We collect a large number of POI search sessions from the search logs of international services at Baidu Maps for offline evaluation. Each example of the dataset consists of the query typed by the user, the POI list that the POI search engine suggested, and the exact POI that the user clicked. Table \ref{table:data_stat} shows the statistics of the large-scale dataset sampled from one-month search logs for model training (abbr. \textit{Train}), hyper-parameter tuning (abbr. \textit{Valid}), and performance testing (abbr. \textit{Test}).

\begin{table}[!htbp]
	\centering
	\caption{
		Statistics of the dataset.
	}
	\begin{tabular}{l|rr}
		\toprule
		\textbf{Subset} & \textbf{ \#(Queries)} & \textbf{\#(Candidate POIs)/\#(Queries)} \\
	    \midrule
		\midrule
		
		\textit{Train}  &       11,935,730     &           2.7                \\
		\textit{Valid}  &        73,255        &           2.8           \\
		\textit{Test}   &      181,589         &          11.6                 \\
		\midrule
		
		\textit{Total}  &       12,190,574     &             2.8             \\
		\bottomrule
	\end{tabular}
	\label{table:data_stat}
\end{table}

\begin{table*}[ht]
	\centering
	\caption{Offline evaluation results of different models.}
	\setlength{\tabcolsep}{4mm}{
		\begin{tabular}{l|rrrrrrr}
			\toprule
			\multirow{2}{*}{\textbf{Model}} & \multicolumn{5}{c}{\textbf{Evaluation Metrics (Offline)}} \\
			& MRR    & nDCG@1    & nDCG@3  & nDCG@10  & SR@1    & SR@3 & SR@10   \\ 
			\midrule
	        \midrule			
			DSSM~\cite{huang2013learning}        & 0.6681 &	0.5235 &	0.6634 &	0.7258 & 	0.5235 &	0.7705 &	0.9324    \\
			ARC-I~\cite{hu2014convolutional}     & 0.6604 &	0.5127 &	0.6561 &	0.7206 & 	0.5127 &	0.7665 &	0.9337    \\
			Conv-DSSM~\cite{shen2014latent}      & 0.6485 &	0.4985 &	0.6420 &	0.7097 & 	0.4985 &	0.7515 &	0.9281    \\
			\midrule
			DPSM~\cite{zhao2019poi}              & 0.6181 &	0.4900 &	0.6084 &	0.6655 & 	0.4900 &	0.7010 &	0.8491    \\ 
			PALM~\cite{zhao2019incorporating}    & 0.6921 &	0.5488 &	0.6902 &	0.7500 & 	0.5488 &	0.7993 &	0.9521    \\
			\midrule
            HGAMN               & 0.7663 &	0.6539 &	0.7653 &	0.8097 & 	0.6539 &	0.8528 &	0.9636    \\
		    w/o POI-POI Graph   &  0.7655 &	0.6527 &	0.7648 &	0.8091 &	0.6527 &	0.8526 &	0.9628     \\
			w/o POI-Query Graph &  0.7573 &	0.6408 &	0.7557 &	0.8030 &	0.6408 &	0.8455 &	0.9640    \\
			w/o Heterogeneous Graph           &  0.6924 &	0.5451 &	0.6921 &	0.7507 &	0.5451 &	0.8052 &	0.9540    \\
			\midrule
			LTR                   & 0.8253 & 0.7323 &	0.8294 &	0.8582 &	0.7323 &	0.9030 &	0.9721       \\
			LTR + HGAMN &  $\mathbf{0.8307}$ & $\mathbf{0.7393}$ &	$\mathbf{0.8347}$ & $\mathbf{0.8627}$ & $\mathbf{0.7393}$ & $\mathbf{0.9072}$ & $\mathbf{0.9743}$  \\ 
			\bottomrule
		\end{tabular}
	}
	\label{table:offline}
\end{table*}

\subsubsection{Evaluation Metrics}
\label{sec:metrics}
We use several widely-used metrics in information retrieval for offline performance evaluation.

The first group of metrics, Success Rate (SR) at Top-K (SR@K), is the coarse metric that denotes the average percentage of ground-truth POIs ranked at or above the position K in the ranked list provided by a POI retriever. Because of the limited space for display on mobile phones, Baidu Maps can mostly display 3 POIs on the first screen when the input keyboard is launched and at most 10 POIs when the input keyboard is closed. Therefore, we consider SR@1, SR@3, and SR@10 for offline evaluation.

Another group of fine-grained metrics, including Mean Reciprocal Rank (MRR) and normalized Discounted Cumulative Gain at Top-K (nDCG@K), concerns more about the exact position where a POI retriever arranges the ground-truth POI in the returned list. We consider nDCG@1, nDCG@3, and nDCG@10 for offline evaluation due to the display limitations on  mobile phones.

\subsubsection{Model Configuration}
The dimensionality of POI and query embeddings $d$ is set to 128. The sequence length of query text, POI name, and POI address is set to 30. The POI graph learning module consists of two graph attention layers with output dimensionality of $d=128$ and $d'=256$, respectively. The number of heads in the multi-head attention $K$ is chosen from $\{1, 2, \ldots, 10 \}$, and finally, set to 4. 

During training, we use Adam optimizer~\cite{kingma2014adam}, with the learning rate initialized to 0.001 and gradually decreased during the process of training. 
To prevent overfitting, we use the dropout strategy with a dropout rate of 0.5. The maximum training epoch is set to 40, and the batch size of the training set is set to 64. 

\subsubsection{Experimental Results}
In this section, we evaluate the effectiveness of HGAMN for the multilingual POI retrieval task. Table~\ref{table:offline} shows the performance of offline assessments on the models mentioned in Section~\ref{sec:model_comp}. From the results, we can see that the proposed model HGAMN significantly outperforms all baseline methods on the large-scale real-world dataset. Specifically, we have the following observations.

(1) HGAMN significantly outperforms all conventional text retrieval methods (i.e., DSSM, ARC-I, and Conv-DSSM). Furthermore, the ``HGAMN w/o POI-POI Graph'' model also achieves better performance compared with these methods. The main reason is that the POI-Query graph is able to model multilingual features between a POI and its historical queries, which enables us to mitigate the gap between a query and the candidate POIs.

(2) Compared with recently proposed neural-based POI retrieval methods (i.e., DPSM and PALM), HGAMN achieves better performance. Although these methods combine geographic or spatial-temporal features with text representations for POI retrieval, they do not take the POI visiting sparsity problem into account, which is a critical problem in industrial map services. The POI-POI graph builds connections between low-frequency POIs and their similar high-frequency ones, which is able to transfer the abundant supervisory signals from high-frequency POIs to facilitate learning better representations of the low-frequency POIs. The  results verify that HGAMN is able to effectively relieve this problem.

(3) After removing the POI-POI graph and POI-Query graph separately (``HGAMN w/o POI-POI Graph'' and ``HGAMN w/o POI-Query Graph''), the performance of HGAMN decays considerably compared with the complete model. This indicates that both components in HGAMN are essential for multilingual POI retrieval, and they are complementary to each other.

(4) In the last section of Table~\ref{table:offline}, we observe that the LTR model outperforms the single HGAMN model. The reason is that the LTR model is one of the typical industrial ranking models based on a large set of time-proven high-quality features~\cite{pmlr-v14-chapelle11a}. 
However, after adding the feature computed by HGAMN into the LTR model, the ``LTR + HGAMN'' model achieves significant improvements. Since it is challenging to create a new feature that is able to significantly improve the overall performance of industrial ranking models, the improvements made by ``LTR + HGAMN'' further confirm the effectiveness of HGAMN. This shows that HGAMN can not only be used as an individual ranking model, but also be used to obtain a single strong feature that is robust to an industrial ranking framework.

\subsection{Online A/B Testing}

\subsubsection{Traffic of Data}
Before being launched in production, we would routinely deploy the new model online and make it randomly serve 5\% traffic of the POI search. During the A/B testing period, we monitor the performance of the new model and compare it with the previously deployed models. This period conventionally lasts for at least one week.

\subsubsection{Experimental Results}
We use SR@1, SR@3, and SR@10 as the metrics for online evaluation, which are also adopted by offline evaluation. Table~\ref{table:online} shows the experimental results of the online A/B testing on different models mentioned in Section~\ref{sec:model_comp}. All models were selected by the test set for offline evaluations, and we launched the best-performed ones. They are tested by 5\% search traffic of Baidu Maps.

\begin{table}[!htbp]
	\centering
	\caption{Online evaluation results of different models.}
	\setlength{\tabcolsep}{4mm}{
		\begin{tabular}{l|rrr}
			\toprule
			\multirow{2}{*}{\textbf{Model}} & \multicolumn{3}{c}{\textbf{Evaluation Metrics (Online)}} \\
			& SR@1    & SR@3  & SR@10  \\ 
			\midrule
			\midrule
			DSSM       &   0.4847 &  0.7130  &  0.8358      \\
			ARC-I      &   0.4668 &  0.7024  &  0.8349      \\
			PALM       &   0.4900 &  0.7010  &  0.8491      \\
			LTR &   0.6647 &  0.8189  &  0.8802      \\
			\midrule
			LTR + HGAMN    &   $\mathbf{0.7173}$ &	 $\mathbf{0.8807}$ &	$\mathbf{0.9437}$      \\
			\bottomrule
		\end{tabular}
	}
	\label{table:online}
\end{table}

Compared with the offline evaluation results, we gain lower results on SR@1, SR@3, and SR@10 in the online A/B testing. The main reason is that the POI lists returned by our POI search engine might be ignored entirely by a small proportion of users, mainly because they prefer directly typing in the full names of their desired POIs and then click the search button. In this interactive mode, Baidu Maps will directly provide users with the relevant POIs. This results in a phenomenon that none of the returned POIs were clicked, which may lead to much lower performance on Success Rate (SR). However, the relative improvements of these models are consistent with those obtained by the offline evaluation.

\section{Discussion}  \label{Discussion}
Here we explore the reason why HGAMN is able to boost both the offline and online performance of the POI search engine. Generally speaking, as an end-to-end framework for multilingual POI retrieval, HGAMN can produce an intermediate feature vector, i.e., the graph-based representation of a POI (denoted by ``HGAMN''). The probability from the classifier, taking the vector as input, is a reliable indicator to decide the rank order of candidate POIs in the model. The significance of this indicator has already been proved by the experimental results of both offline and online evaluations, which are reported by Table~\ref{table:offline} and Table~\ref{table:online}, respectively. 

Moreover, we are curious about how much this probability from the intermediate vector, as a feature, can contribute to the GBRank-based multilingual POI retrieval model LTR, which has kept serving online in the search engine of international service at Baidu Maps. From the perspective of industrial practice, we need to figure out the relative importance of a proposed feature among all features leveraged by the GBRank model for multilingual POI retrieval.

\begin{figure}[ht!]
	\centering
	\includegraphics[width=1.0\linewidth,trim={0.1cm 0.3cm 0.2cm 0.3cm},clip]{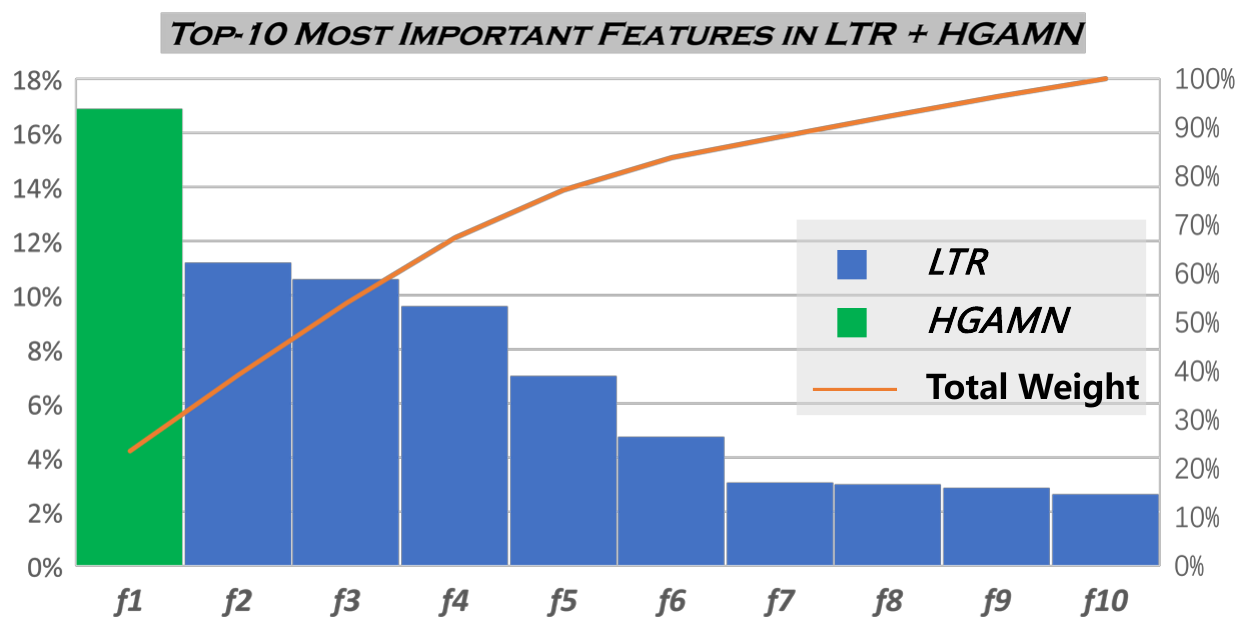}
	\caption{
    	Histogram of the feature importance.
	}
	\label{fig:allfeatures}
\end{figure}

GBRank can provide a score that indicates how useful a feature was in constructing the boosted decision trees within the model, which can help investigate the impact of different features (e.g., \cite{10.1145/3185663,huang2020personalized}). The more a feature is used to make critical decisions for decision trees, the higher its relative importance is allocated. 
Figure~\ref{fig:allfeatures} illustrates the weights of the top-10 most important features in ``LTR + HGAMN'', and the total weight of them is 71.67\%. 
Among all features, the importance of the graph-based representation (i.e., $f1$, colored in green) ranks 1$^{st}$ with the weight of 16.89\%.
This further demonstrates that HGAMN is able to significantly improve the effectiveness of multilingual POI retrieval.

\section{Related Work} \label{related_work}
Here we briefly review the closely related work in the fields of text retrieval, POI retrieval, and heterogeneous graph neural network.

\subsection{Text Retrieval}  
Text retrieval aims to provide the most relevant documents for a query~\cite{yin2016ranking}. There are three conventional categories of methods for text retrieval: pointwise (such as logistic regression~\cite{joachims2002optimizing}), pairwise (such as RankSVM~\cite{joachims2002optimizing} and RankBoost~\cite{Freund1998AnEB}), and listwise (such as ListNet~\cite{cao2007learning} and  AdaRank~\cite{Xu2007AdaRankAB}). The major difference between them lies in the input document space, output space, and loss function. These methods require manually designed features. However, such features may be sparse and insufficient to effectively encode the semantic information of queries and documents. Moreover, designing effective features is usually time-consuming and heavily relies on expert knowledge in particular areas~\cite{Guo2016ADR}.

With the rapid development of deep learning, researchers adopt neural networks to automatically learn representations for text retrieval. For example, \citeauthor{huang2013learning}~\shortcite{huang2013learning} propose a DSSM model to map the query and the document into a semantic space and treat the similarity between two embeddings as the relevance score. Subsequently, Conv-DSSM~\cite{Shen2014LearningSR} and LSTM-DSSM~\cite{palangi2014semantic} are proposed to improve the ability of semantic feature extraction of DSSM. \citeauthor{pang2016text}~\shortcite{pang2016text} propose to model text matching as the problem of image recognition and employ a convolutional neural network to extract matching features. DeepRank~\cite{Pang2017DeepRankAN} further simulates the human judgment process to capture important features. 

Multilingual POI retrieval task is different from text retrieval task in that it requires not only capturing semantic similarities between text data but also addressing the cross-language matching problem and textual-geographic matching problem (i.e., computing the relevance between a query and a POI by taking both text data and geolocations into consideration), which are generally not required in text retrieval task.

\subsection{POI Retrieval}
There is a growing body of work that explores and assesses POI retrieval~\cite{zhao2019incorporating,zhao2019poi,huang2020personalized,li2020personalized,yuan2020spatio,p4ac-kdd21}. Here, we briefly review recent attempts on applying neural networks to address this task. To address the mistyping or an alias inquiry problem, \citeauthor{zhao2019poi}~\shortcite{zhao2019poi} propose a POI latent semantic model based on deep learning, which can effectively extract query and POI features for similarity calculation. \citeauthor{p4ac-kdd21}~\shortcite{p4ac-kdd21} propose a personalized POI retrieval model which also has the ability to provide time- and geography-aware results. Furthermore, geographic information~\cite{zhao2019incorporating} and spatial-temporal factors~\cite{yuan2020spatio} have been considered in recent work. 

However, little work has considered the problems of visiting sparsity and multilingual query-POI matching, which are two main challenges that must be tackled for POI retrieval in global map applications such as Baidu Maps. To address both problems, we first encode a POI's multi-source information to enrich its representation. Then, we establish the relations among different POIs and queries by constructing a heterogeneous graph. Finally, we produce enhanced representations of queries and POIs via the heterogeneous graph, which has a significant effect on POI retrieval performance.

\subsection{Heterogeneous Graph Neural Network}
The heterogeneous graph, which constitutes multiple types of nodes or edges, is ubiquitous in real-world applications.     Previous studies~\cite{sun2018joint,wang2019heterogeneous,cen2019representation,zhu2020hgcn} focus on different aspects of the heterogeneous graph to learn node representations. For example, \citeauthor{sun2018joint}~\shortcite{sun2018joint} propose a meta-graph-based network embedding model, which simultaneously considers the hidden relations of all meta information of a meta-graph. \citeauthor{wang2019heterogeneous}~\shortcite{wang2019heterogeneous} propose a heterogeneous graph neural network, which utilizes hierarchical attention, including node-level and semantic-level attentions, to learn node representations from meta-path based neighbors. \citeauthor{cen2019representation}~\shortcite{cen2019representation} propose a unified attributed multiplex heterogeneous network to solve the multiplex heterogeneous graph embedding problem with both transductive and inductive settings. \citeauthor{zhu2020hgcn}~\shortcite{zhu2020hgcn} propose a heterogeneous graph convolution network to directly learn the complex relational hierarchy, potential incompatible semantics, and node-context relational semantics.

Inspired by the recent success in heterogeneous graph representation learning, we build a heterogeneous graph from search logs, which links the low-frequency POIs with the high-frequency ones and aggregates queries in different languages and formulations for individual POIs. As a result, the visiting sparsity and multilingual matching problem can be effectively alleviated by enhancing the representations of queries and POIs via the heterogeneous graph.

\section{Conclusions and Future Work}  \label{conclusion_and_future}
This paper presents an industrial solution to the multilingual POI search for international services at Baidu Maps. We propose a heterogeneous graph attention matching network (HGAMN) to address the visiting sparsity and multilingual query-POI matching problems. HGAMN is composed of three modules: (1) a multi-source information learning module, which learns the text and location representations of the multilingual query,  POI name, and POI address; (2) a heterogeneous graph learning module, which constructs the connections of different POIs and historical queries, and learns the node representations from the heterogeneous graph; and (3) a POI ranker module, which calculates the relevance between a query and candidate POIs.
We conduct both offline and online evaluations using large-scale real-world datasets. The experimental results show that HGAMN achieves significant improvements over several mainstream approaches, which demonstrates the effectiveness of enhancing the representations of queries and POIs via the heterogeneous graph to improve multilingual POI retrieval.

The user input habits and preferences are not taken into account in this paper. In the future, we intend to utilize these kinds of vital information for personalized POI searches. In addition, previous studies have shown that context \cite{huang2018improving,10.1145/3396501} and explanation \cite{huang2016generating,huang2017learning,huang2018entity} can bring significant improvements in recommendation effectiveness and increase user satisfaction. As future work, we plan to investigate whether multilingual POI retrieval could benefit from the adoption of such factors.

\balance
\bibliographystyle{ACM-Reference-Format}
\bibliography{main}

%%
%% If your work has an appendix, this is the place to put it.
% \appendix

\end{document}